\documentclass[prd,aps,preprint,preprintnumbers,nofootinbib]{revtex4}
\usepackage{epsfig,footnote}
\usepackage{ulem}
\usepackage{color}
\usepackage{array}
\usepackage{amssymb}
\usepackage{amsmath}
\usepackage{graphicx,subfigure}
\usepackage{longtable}
\usepackage{verbatim}
\usepackage{amsfonts}
\usepackage{hyperref}
\usepackage{cancel}
\usepackage{bm}
\newcommand{\degree}{^\circ}
\begin{document}
\title{Analysis of three-body charmed $B$ meson decays $B \to {D}(V^* \to){V P}$}
\author{Jing Ou-Yang, Run-Hui Li, Si-Hong Zhou\footnote{corresponding author: shzhou@imu.edu.cn}}

\affiliation{Inner Mongolia Key Laboratory of 
Microscale Physics and Atom Innovation, 
School of Physical Science and Technology, 
Inner Mongolia University, Hohhot 010021, China}
\affiliation{Center for Quantum Physics and Technologies, 
School of Physical Science and Technology, 
Inner Mongolia University, Hohhot 010021, China}
\date{\today}
\begin{abstract}
We systematically analyze the decays $B_{(s)} \to D_{(s)} (V^* \to)\, V\, P$, 
where $V^*$ represents a vector resonance ($\rho, \, \omega$
or $K^*$), and $V P$ denotes the final-state meson pairs 
$ \omega\,  \pi$, $ \rho\,  \pi$ and $ \rho \, K$.  
The intermediate subprocesses $B_{(s)} \to D_{(s)} V^*$ 
are calculated in the factorization-assisted topological-amplitude 
approach, while the intermediate resonant states $V^*$ are 
modeled using a relativistic Breit-Wigner distribution, 
subsequently decaying into $VP$ through strong interactions. 
We predict the off-shell effects of the ground-state resonances
($\rho,\, \omega, \, K^*$) in $B_{(s)} \to D_{(s)} (V^* \to )V P$.
Our results show that the virtual contributions from
$\rho \to \omega \, \pi$, $\omega \to \rho\, \pi$, and 
$K^* \to \rho\,  K$ are crucial for these three-body decays, 
$B_{(s)} \to D_{(s)} V\, P$. In particular, the branching fractions 
arising from the $\rho$ and $\omega$ virtual effects can be 
comparable to the total decay rates of 
$B_{(s)} \to D_{(s)}  \omega\, \pi$ and 
$B_{(s)} \to D_{(s)}   \rho\, \pi$, respectively.
Decays with branching fractions of order 
$10^{-6}-10^{-4}$ are expected to be measurable
at Belle II and LHCb.
Compared with previous perturbative QCD predictions for
$B_{(s)} \to D_{(s)} (\rho \to)\, \omega\, \pi$,
our results are consistent but exhibit higher precision.
 
\end{abstract}

\maketitle

\section{Introduction}\label{Introduction}
Three-body nonleptonic $B$ meson decays broaden the study 
of $B$ decay mechanisms, enabling tests of the standard model 
(SM) and investigations into the emergence of quantum chromodynamics. 
These decays also provide additional possibilities for {\it CP} violation 
searches through interference between different resonant states,
complementing tree and penguin amplitude interference observed
in two-body $B$ decays. The complex kinematics and phase space 
distributions in three-body decays are typically analyzed using 
Dalitz plot technique~ \cite{Dalitz:1954cq}.
In the edges of the Dalitz plot, the invariant mass squared 
of two final-state particles often peaks, revealing intermediate 
resonances.
These resonance structures provide valuable opportunities to
explore the spectroscopic properties of resonant systems.

Experimentally,  the branching fractions and  {\it CP} 
asymmetries of three-body $B$ meson decays have been
precisely measured by BABAR, Belle (II) and LHCb collaborations
~\cite{Belle:2006wbx,BaBar:2009pnd,BaBar:2010rll,LHCb:2014ioa,
LHCb:2015klp,LHCb:2015tsv,LHCb:2018oeb,Belle-II:2023gye}.
Through amplitude analysis in isobar model,
the fit fractions of both resonant and nonresonant components
can be determined. In this framework, resonances are typically 
parameterized using the relativistic Breit-Wigner (RBW) lineshape,
while nonresonant contributions are described by exponential distributions.
This analysis provides crucial information on their corresponding ground 
and excited resonant states, including their masses, spin-parity 
quantum numbers, and other spectroscopic properties.

However, theoretical frameworks remain challenged by 
nonfactorizable multibody correlations, such as 
final-state rescattering effects~\cite{Bediaga:2015mia,Pelaez:2018qny},
 three-body effects in final states~\cite{Guimaraes:2010zz,Magalhaes:2011sh},
and also the complex interplay between continuum and 
resonant states in three-body phase space.
In the preliminary study of three-body effects,
the heavy meson chiral perturbation 
theory \cite{Cheng:2002qu,Cheng:2007si,Cheng:2013dua} 
have been applied to calculate the nonresonant contributions
in three-body charmless $B$ meson decays, such as 
$B \to KKK, \, B \to K \pi \pi$ processes that are dominated by 
nonresonant components~\cite{Cheng:2008vy}. 
Actually, theoretical interest primarily focus on the resonant 
components of three-body $B$ decays, where two final particles 
from an intermediate resonance that recoils against the 
third ``bachelor"  meson. This configuration, known as 
quasi-two-body decay, benefits from the large energy release 
in $B$ meson decays, where the energetic resonant meson pair 
and bachelor meson move fast back-to-back in the
$B$ meson rest frame.
This kinematic configuration naturally suppresses interactions 
between the meson pair and bachelor particle through 
power suppression, analogous to the  ``color transparency" 
phenomenon observed in two-body $B$ meson decays. 
This suppression enables factorization-based approaches 
for quasi-two-body decays, including 
the QCD Factorization (QCDF) 
\cite{Cheng:2002qu, Cheng:2007si, Cheng:2013dua,Huber:2020pqb,Yuan:2024jvq}, 
the PQCD approach\cite{Chen:2002th,Wang:2014ira,Wang:2016rlo, 
Li:2016tpn, Li:2018psm, Wang:2020plx, Fan:2020gvr, Wang:2020nel, 
Zou:2020atb, Zou:2020fax,Zou:2020ool,Yang:2021zcx,Liu:2021sdw,
Zhang:2022pfn,Zhang:2023uoy,Zhao:2023dnz,Chang:2024qxl} and 
factorization-assisted topological-amplitude (FAT) approach
\cite{Zhou:2021yys,Zhou:2023lbc,Zhou:2024qmm,Zhou:2025nao,Wang:2025rkr}.

In this work, we concentrate on three-body charmed $B$ decays 
 $B_{(s)} \to D_{(s)} (V^* \to )V P$, where $V^*$ represents the 
ground-state resonances $\rho$, $\omega$ and $K^*$, 
 and $V P$ meson pair, originating from $\rho$, $\omega$  and $K^*$ 
 intermediate resonances, denotes the $ \omega\,  \pi$, $ \rho\,  \pi$ 
 and $ \rho \, K$ pairs, respectively. 
Since the pole mass of $\rho(770)$ ($\omega(782)$, $K(892)^*$) 
lies below the invariant mass threshold of the 
$\omega\, \pi$ ($ \rho \, \pi$, $ \rho \, K$) pair, 
the decays $\rho \to  \omega\, \pi^-$ ($\omega \to  \rho\, \pi$, $K^* \to  \rho\, K$)
can only proceed via the off-shell effects~\cite{LHCb:2016lxy} through 
the Breit-Wigner tail (BWT) distribution. 
Nevertheless, this virtual effect in the $B^0 \to \bar D^{(*)-} \omega \pi^-$ 
decay was observed to be significant by Belle, with a branching fraction 
measured as $\mathcal{B} (B^0 \to \bar D^{*-}(\rho(770)^+ \to ) \omega \pi^-) 
 =(1.48^{+0.37}_{-0.63}) \times 10^{-3}$, which is comparable to 
 $\mathcal{B} (B^0 \to \bar D^{*-}(\rho(1450)^+ \to ) \omega \pi^-) 
 =(1.07^{+0.40}_{-0.34}) \times 10^{-3}$~\cite{Belle:2015fvz}.
The off-shell effect of $\rho(770)^+ \to  \omega \pi^-$ also plays
an important role in other processes,
 such as $e^+ e^- \to \omega \, \pi^0$~\cite{BaBar:2017zmc,BESIII:2020xmw,SND:2023gan}
and $\tau \to \omega\, \pi \, \nu_\tau$~\cite{BaBar:2009jyj}.
Although there are no data for any branching fractions of 
$B \to  D (\omega \to ) \rho \pi$ and 
$B \to D (K^* \to ) \rho K$ decays available,
the existing experimental analysis for $\omega \to  \pi^+ \pi^- \pi^0$ 
decay supports the Gell-Mann-Sharp-Wagner
(GSW) mechanism~\cite{Gell-Mann:1962hpq}
that the $\omega \to 3\, \pi$  decay proceeds  
predominantly through intermediate $\omega \to \rho \, \pi$ transition, 
followed by the $\rho \to \pi \pi$ decay.
The GSW mechanism explains the dominance of $\omega \to 3 \pi$
via $\rho\, \pi$ intermediate state, and this likely extends to 
$B \to  D (\omega \to ) \rho\, \pi$ decays, though no direct 
measurements exist.
For the $K^* \to K (\rho \to) \pi \pi $ decays only 
upper limits exist in experimental, theoretical estimates 
would require modeling of the suppressed transitions in 
$B \to D (K^* \to ) \rho K$ firstly.
The theoretical analysis of the process $B^0 \to D^{*-} \, \omega\, \pi$ 
has been done in Refs.~\cite{Matvienko:2011ic,Eidelman:2019mee}
using the factorization hypothesis, but without providing 
observable predictions for its branching fraction. 
Recently, the branching fractions of 
$B_{(s)} \to D_{(s)}^{(*)} (\rho(770,1450)^+ \to)\, \omega\, \pi^+$
have been calculated within the PQCD approach~\cite{Ren:2023kxc}.
Actually, the off-shell effects in $B \to D P P$ through
 $D^* \to D P$ have already been found to be essential 
 for the total amplitudes by Belle~\cite{Belle:2003nsh}, 
 BABAR~\cite{BaBar:2009pnd} and 
 LHCb~\cite{LHCb:2015eqv,LHCb:2015tsv,LHCb:2015klp,LHCb:2015jfh,LHCb:2016lxy},
 and the corresponding studies on the theoretical side
 were performed firstly in the PQCD approach~\cite{Wang:2018dfq,Chai:2021kie},
 followed by more precise results in the FAT approach~\cite{Zhou:2021yys},
which improved upon the PQCD predictions.

For a systematic study of $B_{(s)} \to D_{(s)} (V^* \to )V P$ decays,
we will apply the FAT approach. This framework was originally developed by 
one of us (S.-H. Z.) and collaborators~\cite{Li:2012cfa, Li:2013xsa, Zhou:2015jba, Zhou:2016jkv, 
 Jiang:2017zwr, Zhou:2019crd, Qin:2021tve}
 to address nonfactorizable contributions in two-body 
 $D$ and $B$ meson decays, and 
 has subsequently been successfully extended to both quasi-two-body $B$ 
meson decays~\cite{Zhou:2021yys,Zhou:2023lbc,Zhou:2024qmm} and 
$D$ meson decays~\cite{Zhou:2025nao}.
Building upon the conventional topological diagram approach~\cite{Cheng:2010ry}, 
the FAT approach systematically classifies 
decay amplitudes according to distinct electroweak Feynman diagrams 
but keeping the SU(3) symmetry breaking effects.
Specifically, we incorporate these symmetry-breaking effects
 into topological diagram amplitudes 
primarily through factorizing out form factors and decay constants, 
assisted by QCD factorization. 
The remaining nonfactorizable contributions are 
parameterized as a minimal set of universal parameters
which are determined through a global fits to 
all available experimental data.
This makes the FAT approach particularly powerful for
providing precise decay amplitudes for
intermediate two-body processes $B \to D V^*$ involving 
charmed final states. 
In our analysis of $B_{(s)} \to D_{(s)} (V^* \to )V P$ decays, 
we describe the subsequent strong decays of 
intermediate resonances to final-state meson pairs 
$\omega\, \pi$, $\rho\, \pi$ and $\rho\, K$ in terms of 
RBW formalism, consistent with experimental analyses. 

The remainder of this paper is organized as follows. 
 In Sec.~\ref{sec:2}, we give an introduction of the 
 theoretical framework.  
 Numerical results and detailed discussions about 
 $B_{(s)} \to D_{(s)} (\rho \to ) \omega \pi$,
 $B_{(s)} \to D_{(s)} (\omega \to ) \rho \pi$,
and $B_{(s)} \to D_{(s)} (K^* \to ) \rho K$ 
are collected in Sec.~\ref{sec:3}. 
Sec.~\ref{sec:4} has our conclusions.


\section{Factorization Amplitudes for Topological Diagrams}\label{sec:2}

\begin{figure} [htb]
\begin{center}
\scalebox{1}{\epsfig{file=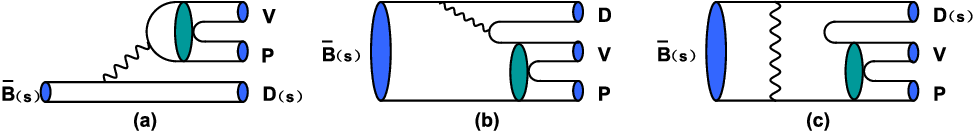}}
\caption{Topological diagrams of $\bar{B}_{(s)}
 \rightarrow{D_{(s)}}(V^* \rightarrow) P V$ under the framework 
 of quasi-two-body decay with the wave line representing a 
 W boson: (a) the color-favored emission diagram $T$, 
 (b) the color-suppressed emission diagram $C$, 
 and (c) the W-exchange diagram $E$.}
\label{btocdiagrams}
\end{center}
\end{figure}

\begin{figure} [htb]
\begin{center}
\scalebox{1}{\epsfig{file=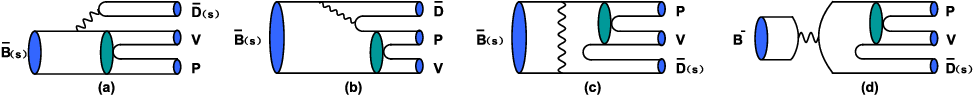}}
\caption{The same as Fig.~\ref{btocdiagrams}, but for 
 $\bar{B}_{(s)} \rightarrow{\bar{D}_{(s)}}(V^* \rightarrow)P V$, 
 with additional W-annihilation diagram $A$ in (d).}
\label{btoudiagrams}
\end{center}
\end{figure}

The charmed quasi-two-body decays 
$B_{(s)}\rightarrow{ D_{(s)}} (V^* \rightarrow)P V$ 
proceed through two distinct stages.
Initially, the $B_{(s)}$ meson decays into $D_{(s)}$ meson and 
an intermediate resonant state $V^*$. 
The  unstable resonance $V^*$ subsequently decays into 
a light pseudoscalar meson $P$ and a vector meson $V$
via strong interactions. At quark level, the 
first subprocess  is induced by weak transitions 
$b \rightarrow c q \bar{u} (q = d, s)$ and 
$b \rightarrow u q \bar{c} (q = d, s)$, leading to 
final-state $D_{(s)} V^*$ and $\bar{D}_{(s)} V^*$, respectively. 
According to the weak interactions of $b \rightarrow c q \bar{u}$, 
the topological diagrams contributing to 
$\bar{B}_{(s)}\rightarrow{ D_{(s)}} (V^* \rightarrow)P V$ 
can be classified into three types as listed in Fig.~\ref{btocdiagrams}, 
(a) color-favored emission diagram $T$, 
(b) color-suppressed emission diagram $C$, 
and (c) W-exchange diagram $E$. 
For $\bar{B}_{(s)}\rightarrow{ \bar{D}_{(s)}} (V^* \rightarrow)P V$,
induced by $b \rightarrow u q \bar{c}$ transitions,
an additional W-annihilation diagram $A$ is included, 
as shown in Fig. \ref{btoudiagrams}.

The amplitudes of the first subprocess, ${B}_{(s)}\rightarrow{ D_{(s)}} V^*$, 
can be related to those of two-body charmed $B$ decays within
the framework of the FAT approach~\cite{Zhou2015}. 
The factorization theorem for the $T$ diagram amplitude 
has been established at high precision by the QCD factorization, 
the perturbative QCD based on $k_T$ factorization, 
as well as the soft-collinear effective theory~\cite{Keum:2003js,Beneke:2000ry,Bauer:2001cu}.
Meanwhile, significant nonfactorizable effects 
have been observed in both the $C$ and $E$ topologies. 
Following the FAT approach~\cite{Zhou2015}, 
we parameterize the matrix elements for these 
nonfactorizable contributions, while adopting 
a well-established factorization formula for the $T$
diagram amplitude, 
expressed as
\begin{align}\label{2bodyamp}
\begin{aligned}
&T^{DV^*}=\sqrt{2}\, G_F\,  V_{cb} V_{uq}^*\,  a_1(\mu)\, f_{V^*} \, m_{V^*}\,  
F_1^{B \rightarrow{D}}(m_{V^*}^2)(\varepsilon_{V^*}^* \cdot p_B),\\
    &C^{DV^*}=\sqrt{2}\, G_F V_{cb} V_{uq}^*\,  f_{D_{(s)}} \, m_{V^*} \,
     A_0^{B\rightarrow{V^*}}(m_D^2)(\varepsilon^*_{V^*} \cdot p_B)\, \chi^C \, e^{i\phi^C},\\
    &E^{DV^*}=\sqrt{2}\, G_F\,  V_{cb} V_{uq}^* \,  m_{V^*} \, 
    f_B\,  \frac{f_{D_{(s)}}f_{V^*}}{f_D f_\pi}(\varepsilon^*_{V^*} \cdot p_B)\,  \chi^E \, e^{i\phi^E}\, ,
\end{aligned}
\end{align}
where $a_1({\mu})$ in $T^{D V^*}$ is the effective Wilson coefficient.
The decay constants $f_{V^*}$, $f_{D_{(s)}}$ and $f_B$ 
correspond to the vector mesons, $D_{(s)}$ mesons 
and $B$ mesons, respectively. 
The  form factors $F_1^{B\rightarrow D}$ and $A_0^{B\rightarrow{V^*}}$
describe the $B_{(s)}\rightarrow D$ and $B_{(s)} \rightarrow V^*$ 
transitions amplitude, respectively.
Here, $\varepsilon_{V^*}$ denotes the polarization vector of $V^*$.
The parameters $\chi^{C(E)}$ and $\phi^{C(E)}$ in $C^{D V^*}(E^{D V^*})$
characterize the magnitude and phase of 
nonfactorizable contributions after factorizing out
decay constants and form factors. 
These unknown parameters are determined through global fits to experimental data. 
For Cabibbo-suppressed $\bar{B}_{(s)} \rightarrow \bar{D}_{(s)} V^*$ decays,
the currently available experimental data remain insufficient to 
perform a global fit for extracting nonfactorizable parameters 
in their $C$ and $E$ amplitudes. 
We consequently use the same nonfactorizable parameters 
$\chi^C$, $\phi^C$, $\chi^E$, $\phi^E$ as in 
$\bar{B}_{(s)} \rightarrow D_{(s)} V^*$ decays, as done in Ref.~\cite{Zhou2015}. 
For the $A$ topology of $\bar{B}_{(s)} \rightarrow \bar{D}_{(s)} V^*$ decays, 
the dominant factorizable contribution admits a pole-model computation~\cite{Zhou2015}, 
with the explicit expression given by
\begin{align}
\begin{aligned}
A^{\bar D V^*}=-\, \sqrt{2}\, G_F\,  V_{ub} \, V_{cq}^* \, a_1(\mu)\, 
f_B \, \frac{f_{D_{(s)}}\,  g_{D D V^*}\, m_D^2}{m_B^2\, -\, m_D^2}(\varepsilon^*_{V^*} \cdot p_B)\, ,
\end{aligned}
\end{align}
where $g_{D D V^*} = 2.52$ is the effective strong coupling constant, 
derived from the vector meson dominance model \cite{Lin2000}.

In the following, we will demonstrate the calculation of 
the subsequent subprocess, e.g., the decay of intermediate 
resonances into final states through strong interactions, 
$V^*\rightarrow PV$. 
The related effective Lagrangian, with strong coupling coefficient
 $g_{V^*V P}$, is written as~\cite{Boal:1976eg,Kaymakcalan:1983qq,Fajfer:1992hi},
\begin{eqnarray}   
     \mathcal{L}_{V^*V P}
            = g_{V^*V P} \, \epsilon_{\mu\nu\alpha\beta}\, 
            \partial^\mu V^{*\nu}  \,  \partial^\alpha V^\beta P\, ,
  \label{strongLag}  
\end{eqnarray}
and the matrix element~\cite{Pacetti:2008hd,Caprini:2015wja,Schafer:2023qtl}
\begin{eqnarray}   
      \langle V (p_V,\lambda) P(p_P)| j_\mu(0) | 0 \rangle      
            = i  \epsilon_{\mu\nu\alpha\beta}\, \varepsilon^{\nu}(p_V, \lambda)\, p_P^\alpha \, p^\beta F_{VP}(s)\, ,
  \label{strongmatrix}  
\end{eqnarray}
where $j_\mu$ is the isovector part of the electromagnetic current,
$\lambda$ is the polarization parameter.
In this expression, the form factor $F_{VP}(s)$, describing 
$V^*\rightarrow PV$ transition, is parametrized 
in the vector meson dominance model 
as~\cite{Achasov:2013btb,Achasov:2000wy,Achasov:2016zvn,Fang:2021wes}
\begin{eqnarray}
	F_{VP}(s) = \frac{g_{V^*V P}}{f_{V^*}}
	               \frac{m_{V^*}^2 }{D_{V^*}(s)} \, .
  \label{exp-formfactor}
\end{eqnarray}   
We utilize the RBW distribution for the resonance $V^*$, 
as used in experiments~\cite{LHCb2019,BaBar2011,BaBar2012}. 
The denominator $D_{V^*}(s)$ in Eq.(\ref{exp-formfactor}) is given by
\begin{equation}
    D_{V^*}(s)=m_{V^*}^2\, -\,  s\,-\, i \, \sqrt {s} \, \Gamma_{V^*} (s),
\end{equation}
 where $s=(p_P +p_V)^2$ 
represents the invariant mass square of the pseudoscalar 
and vector mesons with 4-momenta $p_P$, $p_V$. 
The width of vector resonances $\Gamma_{V^*} (s)$ 
depends on the variable $s$ as
\cite{SND:2023gan},
\begin{equation}
\label{Gamma0}
    \Gamma_{V^*} (s)=\Gamma_0 \,  \frac{m_{V^*}^2}{s}\,
     \left(\frac{\mathbf{q}(s)}{\mathbf{q}_0}\right)^3\, 
    +\, 
    \frac{g_{V^*V P}^2}{12 \pi}\, \mathbf{q}^3 (s)\, ,
\end{equation}
where the first term corresponds to  $V^* \to P P$ decays, 
and the second term describes $V ^*\to V P$ decays, 
which are the focus of this work. 
Here, 
\begin{equation}\label{momentumq}
\mathbf{q}(s)=\frac{1}{2 \sqrt{s}}\sqrt{[s-(m_{p_V}+m_{p_P})^2][s-(m_{p_V}-m_{p_P})^2]}\, ,
\end{equation}
represents the momentum magnitude of the final state particles $P$ or $V$ 
in the rest frame of resonance $V^*$. The quantity $\mathbf{q}_0$ denotes
the on-shell value of $\mathbf{q}$ at $s=m_{V^*}^2$. 
In Eq. (\ref{Gamma0}), $\Gamma_0$ represents the full width
of the resonant state, with its value taken from the 
Particle Data Group (PDG)~\cite{ParticleDataGroup:2024cfk}.
The full widths of $\rho, \, \omega,\, K^*$, along with their corresponding resonance masses $m_{V^*}$,
are listed in Table~\ref{tab:mass and width}.
\begin{table}[tbhp]
\caption{Masses $m_V$ and full widths $\Gamma_0$  ($\mathrm{MeV}$) 
of vector resonant states.}
\vspace{3mm}
\label{tab:mass and width}
\centering
\begin{tabular}{cccc}
\hline
Resonance ~~&~~ Line shape Parameters~~ &
~~Resonance ~~&~~ Line shape Parameters
\\ \hline
$\rho(770)~~$  & $m_V\, =\, 775.26 \pm 0.23\,  \, $ &
$\omega(782)~~$  & $m_V\, =\, 782.66\pm0.13\,  \, $ \\
& $\Gamma_0\, =\,  147.4 \pm 0.8\, \,$
& &$\Gamma_0\, =\, 8.68\pm0.13\, \, $\\
$K^*(892)^+~~$  & $m_V\, =\, 891.67\pm0.26\,  \,$ 
&$K^*(892)^0~~$  & $m_V\, =\, 895.55\pm0.20\,  \, $\\
& $\Gamma_0\, =\, 51.4\pm0.8\, \, $
 & &$\Gamma_0\, =\, 47.3\pm0.5\, \, $\\
\hline
\end{tabular}
\end{table}

The strong coupling constant $g_{V^* V P}$, defined in Eq.(\ref{strongLag}),
describes the strong interactions among the three mesons at the hadron level.
Its value in $V^* \to V P$ decays can be estimated using
the relation $g_{V^* V P} \approx 3\, g^2_{V^* P P} /(8 \pi^2\, F_\pi)$~\cite{Boal1976}
with $F_{\pi}= f_\pi/ \sqrt {2}=92$ MeV~\cite{ParticleDataGroup:2024cfk},
and $g_{\rho \pi \pi}=6.0$, $g_{K^* K  \pi}=4.59$~\cite{Zhou:2024qmm}. 
Using these inputs, we obtain
$g_{\rho \omega \pi}=14.8\, \mathrm{GeV}^{-1}$ and $g_{K^* K  \rho}=8.69\,  \mathrm{GeV}^{-1}$.
The coupling $g_{\rho \omega \pi}$ has also been determined 
experimentally by SND Collaboration~\cite{Achasov:2016zvn} yielding 
$15.9 \pm 0.4 \, \mathrm{GeV}^{-1}$ and $16.5 \pm 0.2 \, \mathrm{GeV}^{-1}$,
depending on the form factor model used. 
For the numerical calculation of this work, we adopt 
$g_{\rho \omega \pi} = 16.0 \pm 2.0$\, $\text{GeV}^{-1}$, 
consistent with the value used in Ref.~\cite{Ren_2024},
 to compare with the PQCD's results.
The numerical results of $g_{\omega \rho \pi}$ are cited 
from~\cite{Lublinsky:1996yf} calculated by QCD sum rules, 
$g_{\omega \rho \pi}= 16.0$  $\text{GeV}^{-1}$,
which are expected to be equal to $g_{\rho \omega \pi}$
due to the similarity of the $\rho$ and $\omega$ in the context of this vertex.

Finally, by combining the two subprocesses together, 
one can obtain the decay amplitudes of the topological 
diagrams of $B\rightarrow D(V^* \rightarrow)PV$ shown 
in Fig.\ref{btocdiagrams} and Fig.\ref{btoudiagrams}, which are given as 
follows, 
\begin{align}
\begin{aligned}
\label{tbc}
T &= \langle V(p_V) P(p_P) |(\bar{q}u)_{V-A}|0\rangle \langle D(p_D) | (\bar{c}b)_{V-A} | B(p_B) \rangle\\
    &=\sqrt{2}\, G_F\, V_{cb}V_{uq}^*\,  a_1\,  f_{V^*}\,  m_{V^*} \, F_1^{B\rightarrow{D}}(s)\, 
         \epsilon_{\mu\nu\alpha\beta} \,  p_D^\mu\,  \varepsilon^{\nu}_V \, p_P^\alpha\,  (p_V+p_P)^\beta\, 
          \frac{g_{V^* V P } }{D_{V^*}(s) } \,  \\
C &= \langle  V(p_V) P(p_P) |(\bar{q}b)_{V-A}|B(p_B)\rangle \langle D(p_D) | (\bar{c}u)_{V-A} |  0\rangle\\
   &=\sqrt{2}\, G_F\,  V_{cb}V_{uq}^* \,  f_{D} \, m_{V^*} 
        A_0^{B\rightarrow{V^*}}(m_D^2)\, \chi^C \, \mathrm{e}^{i\phi^C}
         \epsilon_{\mu\nu\alpha\beta} \,  p_D^\mu\,  \varepsilon^{\nu}_V \, p_P^\alpha\,  (p_V+p_P)^\beta\, 
          \frac{g_{V^* V P } }{D_{V^*}(s) } \,  \\  
E &= \langle D(p_D) V(p_V) P(p_P) |\mathcal{H}_{eff}|B(p_B)\rangle \\
      &=\sqrt{2}\, G_F\,  V_{cb}V_{uq}^*\,  m_{V^*}\,  f_B \, \frac{f_{D_{(s)}}f_{V^*}}{f_D f_\pi} \, 
          \chi^E \, \mathrm{e}^{i\phi^E} \,
        \epsilon_{\mu\nu\alpha\beta} \,  p_D^\mu\,  \varepsilon^{\nu}_V \, p_P^\alpha\,  (p_V+p_P)^\beta\,  
          \frac{g_{V^* V P } }{D_{V^*}(s) } \, \\
\end{aligned}
 \end{align}
for $b\rightarrow{c}$ transitions, and
\begin{equation}
\label{t}
    \begin{split}
        T &= \langle V(p_V)  P(p_P) |(\bar{u}b)_{V-A}|B(p_B)\rangle \langle \bar{D}(p_{\bar{D}}) | (\bar{c}q)_{V-A} | 0 \rangle\\
         &=\sqrt{2}\, G_F\,  V_{ub}V_{cq}^* \, a_1\,  f_{D} \, m_{V^*} \, 
          A_0^{B\rightarrow{V^*}}(m_{D}^2)\, 
        \epsilon_{\mu\nu\alpha\beta} \, p^\mu_{\bar{D}}\, \varepsilon^{\nu}_V \, p_P^\alpha\,  (p_V+p_P)^\beta\,
         \frac{g_{V^* V P } }{D_{V^*}(s) }  \\
        C &= \langle P(p_P) V(p_V)|(\bar{u}b)_{V-A}|B(p_B)\rangle \langle \bar{D}(p_{\bar{D}}) | (\bar{c}q)_{V-A} | 0 \rangle\\
         &=\sqrt{2}\, G_F\, V_{ub}V_{cq}^* \,  f_{D} \, m_{V^*} \, 
          A_0^{B\rightarrow{V^*}}(m_{D}^2)\, \chi^C\,  \mathrm{e}^{i\phi^C}\, 
        \epsilon_{\mu\nu\alpha\beta} \, p^\mu_{\bar{D}}\, \varepsilon^{\nu}_V \, p_P^\alpha\,  (p_V+p_P)^\beta\, 
         \frac{g_{V^* V P } }{D_{V^*}(s) } \\
        E &= \langle \bar{D}(p_{\bar{D}}) P(p_P) V(p_V)|\mathcal{H}_{eff}|B(p_B)\rangle \\
         &=\sqrt{2}\, G_F\,  V_{ub}V_{cq}^* \,  m_{V^*} \, f_B \, \frac{f_{D_{(s)}}f_{V^*}}{f_D f_\pi} \,
         \chi^E\, \mathrm{e}^{i\phi^E}\, 
        \epsilon_{\mu\nu\alpha\beta} \, p^\mu_{\bar{D}}\, \varepsilon^{\nu}_V \, p_P^\alpha\,  (p_V+p_P)^\beta\,  
         \frac{g_{V^* V P } }{D_{V^*}(s) }\\
        A&= \langle \bar{D}(p_{\bar{D}}) P(p_P) V(p_V)|\mathcal{H}_{eff}|B(p_B)\rangle \\
         &=\sqrt{2}\, G_F\,  V_{ub}V_{cq}^*  \, a_1\,  f_B\, \frac{f_D g_{D D V^* }m_D^2}{m_B^2-m_D^2} \, 
        \epsilon_{\mu\nu\alpha\beta} \, p^\mu_{\bar{D}}\, \varepsilon^{\nu}_V \, p_P^\alpha\,  (p_V+p_P)^\beta\,
         \frac{g_{V^* V P } }{D_{V^*}(s) }  \\
    \end{split}
\end{equation}
for $b \rightarrow{u}$ transitions.
Here the form factor $F_1^{B \to D}(s)$ varies dynamically with 
the invariant mass squared $s$ of $VP$ system, 
unlike that in two-body decays fixed at $s=m_{V^*}^2$.
The decay amplitudes of $B\rightarrow{DVP}$ in Eq.(\ref{tbc}) 
and Eq.(\ref{t}) can be expressed as 
\begin{equation}
    \langle D(p_D)  V(p_V) P(p_P)|\mathcal{H}_{eff}|B(p_B)\rangle
    \, =\,  
        \epsilon_{\mu\nu\alpha\beta} \,  p_D^\mu\,  \varepsilon^{\nu}_V \,
         p_P^\alpha\,  (p_V+p_P)^\beta\, \mathcal{A}(s)\, ,
\end{equation}
where $\mathcal{A}(s)$ represents the sub-amplitude from Eq.(\ref{tbc}) and Eq.(\ref{t}). 
Using the polarization sum relation,
\begin{eqnarray}
 &&\sum_{\lambda=0,\pm}\vert
        \epsilon_{\mu\nu\alpha\beta} \,  p_D^\mu\,  \varepsilon^{\nu}_V (p_V,\lambda)\, 
        p_P^\alpha\,  (p_V+p_P)^\beta\, \vert^2 
= s\,\vert \mathbf{p}_V\vert^2 \vert \mathbf{p}_D \vert^2 (1-\cos^2{\theta})\, ,
\end{eqnarray}
we derive the differential branching fraction for $B \rightarrow{ D V P}$, and the result is
\begin{equation}\label{dGamma}
  \frac{ \mathrm{d}\mathcal{B}}{\mathrm{d}s}
 = \frac{s\, |\mathbf{p}_V |^3 |\mathbf{p}_{D}|^3}{96 \pi^3\, m_B^3}\,  \mathcal{A}(s)^2\, \tau_B\,  ,
\end{equation}
where $|\bm{p}_D|$ and $|\bm{p}_V|$ are the magnitudes of 
 the 3-momenta  $p_D$ and $p_V$ in the rest frame of the resonance, 
  given by
\begin{equation}
 \begin{split}
      |\bm{p}_D|&=\frac{1}{2\sqrt{s}}\sqrt{(m_B^2-m_D^2)^2-2(m_B^2+m_D^2)s+s^2},  
 \end{split}
\end{equation}
and $ |\mathbf{p}_V|=q$ [see Eq.(\ref{momentumq})]. 
The quantity $\tau_B$ is the lifetime of the $B$ meson.

 
\section{Numerical Results and Discussion}\label{sec:3}

\subsection{Input parameters}
The input parameters are classified into three categories: 

(i) Electroweak coefficients:  These contain CKM matrix elements 
and effective Wilson coefficients. The CKM matrix is parameterized 
in the Wolfenstein scheme, with parameters from PDG~\cite{ParticleDataGroup:2024cfk}
\begin{equation}
    \lambda = 0.22501 \pm 0.00068,\,\ A = 0.826^{+ 0.016}_{-0.015},\,
    \ \bar{\rho}= 0.1591\pm 0.0094, \,\ \bar{\eta}= 0.3523 ^{+0.0073}_{-0.0071}.
\end{equation}
The effective Wilson coefficient $ a_1(\mu)=C_2(\mu)+C_1/3 $ is 
calculated as 1.036 at the scale $\mu=m_b/2$.

(ii)  Hadronic parameters: The resonance mass $m_{V^*}$ 
and its full width $\Gamma_0$ was listed in Table~\ref{tab:mass and width}. 
The strong coupling constant $g_{V^* V P}$ involved in the resonance 
$V^*$ decays was also introduced in the preceding section.

(iii) Nonperturbative QCD parameters: These include decay constants, 
transition form factors, and nonfactorizable parameters $\chi^{C(E)}$ 
and $\phi^{C(E)}$.
The decay constants of pseudoscalar mesons ($B, \, B_s, \, D, \, D_s $) 
and light vector mesons ($\rho, \, K^*$), 
as well as the transition form factors $F_1 (Q^2)$ and $A_0 (Q^2)$ for 
$B$ meson decays at zero recoil ($Q^2 =0$), 
are listed in Tables~\ref{tab:decay constants} and 
~\ref{transition form factors}, respectively. 
\begin{table}[tbhp]
\caption{ The decay constants of mesons (in unit of MeV). }
\vspace{3mm}
\label{tab:decay constants}
\newsavebox{\tablebox}
\begin{lrbox}{\tablebox}
\centering
\begin{tabular}{ccccccccccc|}
\hline
~~~~$f_{B}$~~~~ & ~~~~$f_{B_s}$~~~~  & ~~~~ $f_{D}$ ~~~~& ~~~~
$f_{D_s}$~~~~ &~~~~$f_{\rho}$ ~~~~&~~~~$f_{\omega}$ ~~~~&~~~~$f_{K^*}$ ~~~~&
\\ \hline
$190.0 \pm 1.3$ ~~& $230.3 \pm 1.3 $ ~~& $212.0 \pm 7.0 $ ~~& $249.9 \pm 5.0 $ ~~& $213 \pm 11$ ~~& $192 \pm 10$~~ & $220 \pm 11$ &
\\
\hline
\end{tabular}
\end{lrbox}
 \scalebox{0.9}{\usebox{\tablebox}}
\end{table}
\begin{table} [hbt]
\caption{The transition form factors and dipole model parameters. }\label{transition form factors}
\vspace{3mm}
\centering
\begin{tabular}{|c|c|c|c|c|c|c|c|c|c|c|}
\hline&
$~~F_{1}^{B\to D}~~$&
$~~F_{1}^{B_s\to D_s}~~$&
$~~A_{0}^{B\to \rho}~~$&
$~~A_{0}^{B\to \omega}~~$&
$~~A_{0}^{B\to K^*}~~$&
$~~A_{0}^{B_s\to K^*}~~$\\
\hline
$~F_i(0)~$&
0.54&
0.58&
0.30 &
0.26&
0.33&
0.27\\
$\alpha_1$&
2.44&
2.44&
1.56&
1.60&
1.51&
1.74\\
$\alpha_2$&
1.49&
1.70&
0.17&
0.22&
0.14&
0.47\\
\hline
\end{tabular}
\end{table}
The decay constants of $B_{(s)}$ and $D_{(s)}$ are derived 
from a global fit to experimental data by the PDG~\cite{ParticleDataGroup:2024cfk}. 
The decay constants of  the vector 
mesons $\rho$, $\omega$ and $K^*$, as well as all form factors, 
are obtained only using various theoretical approaches, 
such as light-cone sum rules \cite{Wang2015,Gao2020}. 
We utilize the same theoretical values as in our previous 
work on $B \to D (V \to) P P$~\cite{Zhou:2024qmm}, 
assigning a 5\% uncertainty to these decay constants and 
a 10\% uncertainty to the form factors $F_1 (0)$ and $A_0 (0)$. 
The $Q^2$-dependence of form factors can be described
using various parametrizations, such as the $z$-series 
parametrization, applied to $B \to \pi ,K $~\cite{Cui:2022zwm}, 
$B \to \rho, \omega, K^{*} $~\cite{Gao:2019lta}
and $B_{(s)} \to D_{(s)}^{(*)} $ \cite{Cui:2023jiw},
and also pole model parametrization, widely used for 
 $B_{(s)} \to V $ and $B_{(s)} \to D_{(s)}$ transitions.
 For (quasi-)two-body $B$ decays, we focus on the form factors
at a fixed kinematic point ($Q^2=0$), where the choice of 
parametrization has negligible impact.
Following the Refs.~\cite{Zhou2015} and~\cite{Zhou:2024qmm},
we apply the dipole model,
 \begin{equation}\label{eq:ffdipole}
F_{i}(Q^{2})={F_{i}(0)\over 1-\alpha_{1}{Q^{2}\over m_{\rm pole}^{2}}
+\alpha_{2}{Q^{4}\over m_{\rm pole}^{4}}}\, ,
\end{equation}
where $F_{i}$ denotes $F_{0,1}$ or $A_{0}$, and $m_{\rm pole}$ 
is the mass of the corresponding pole state ( e.g.,$B^{*}$ for 
$F_{0,1}$ and $B$ for $A_{0}$). The dipole parameters $\alpha_{1,2}$ 
are given in Table~\ref{transition form factors}.
The nonfactorizable parameters $\chi^{C(E)}$ and $\phi^{C(E)}$ 
 in Eq.(\ref{2bodyamp}) were fitted with updated experimental data 
 from PDG~\cite{ParticleDataGroup:2024cfk} in Ref~\cite{Zhou:2024qmm}, 
 yielding 
\begin{align}
\begin{aligned}
\label{fitpara}
\chi^{C}&=0.453 \pm 0.007, ~~~~~ 
\phi^{C}=\left(65.06\pm 0.83\right)^{\degree}, \\
 \chi^{E}&=0.023\pm0.001,~~~~~ 
 \phi^{E}=\left(142.54\pm 5.99\right)^{\degree}\, .
 \end{aligned}
\end{align}
with $\chi^{2}/\mathrm{d.o.f.}=1.6$.

By incorporating the three categories of input parameters 
into Eq.(\ref{dGamma}), we integrate the differential 
branching fraction over variable $s$ 
in the whole accessible kinematic range to 
obtain the branching fractions for 
$B \rightarrow D(V^* \rightarrow) V P$ decays. 
Our analysis concentrate on three specific channels:
$\bar{B}_{(s)} \rightarrow D_{(s)} (\rho \rightarrow) \omega \pi $,
$\bar{B}_{(s)} \rightarrow D_{(s)} (\omega \rightarrow) \rho \pi$
 and 
$\bar{B}_{(s)} \rightarrow D_{(s)} (K^* \rightarrow) \rho K$, 
along with their doubly CKM-suppressed counterparts
$ \bar{B} \rightarrow \bar{D}(V^* \rightarrow) V P$.
The numerical results are collected in 
Tables~\ref{rhoVP} $-$ \ref{KstarrhoK}, respectively. 
The uncertainties in our predictions ($\mathcal{B}_{\text{FAT}}$) 
arise sequentially from the fitted parameters, the form factors and
the decay constants for $\bar{B} \rightarrow D(V^* \rightarrow) V P$, 
and an additional uncertainty from $V_{ub}$ for $b\rightarrow u$ 
induced $\bar{B} \rightarrow \bar{D}(V^* \rightarrow) PV$ decays. 
The precision of these predictions can be further improved with
more precise determinations of the form factors.
For clarity, the three tables include the CKM matrix elements,
the intermediate resonance decays, as well as the 
topological diagram amplitudes abbreviated as $T$, $C$, $E$, and $A$
with prefactors $1,\, -1,\,  \pm1/\sqrt{2 }$) for the sake of isospin
and flavor mixing.
The PQCD predictions are also listed in the last column of Table~\ref{rhoVP} 
for comparison.



\subsection{Off-shell effects of $\rho$, $\omega$ and $K^*$ in $ B_{(s)} \to D \, V\, P$ decays}
\begin{table}[!hb]
\caption{Branching fractions in the FAT approach of quasi-two-body decays
$\bar B_{(s)} \to D_{(s)} (\rho\to) \omega \pi$ (top) with the uncertainties 
from the fitted parameters, form factors and decay constants, respectively,
and $\bar B_{(s)} \to \bar D_{(s)} (\rho \to) \omega \pi$ (bottom)
with an additional uncertainty from $V_{ub}$ in the last one.
The  PQCD predictions are also list in last column for comparison.       
The CKM matrix elements and topological diagram contributions denoted by
$T$, $C$, $E$ and $A$ are listed in the second column.}
 \label{rhoVP}
\begin{center}
\begin{tabular}{cccc}
 \hline \hline
~~~~{Decay Modes}~~~~     &  ~~~~Amplitudes~~~~ &  ~~~~$\mathcal{B}_{\text{FAT}}$ ~~~~   
 & ~~~~ $\mathcal{B}_{\text{PQCD}}$ \cite{Ren_2024} ~~~~  \\
\hline
$\bar{B} \rightarrow{D}(\rho\rightarrow){\omega \pi}$ & $V_{cb}V_{ud}^*$ &$(\times 10^{-3})$ & $(\times 10^{-3})$\\
$B^- \to D^0(\rho^-\to) \omega \pi^-$ & $T+C$ & $0.80^{+0.01+0.15+0.07}_{-0.01-0.13-0.07} $ &$1.42^{+0.16+0.15+0.11+0.10}_{-0.16-0.13-0.09-0.10}$  \\
                           
$\bar{B}^0 \to D^+ (\rho^-\to) \omega \pi^-$ &$T+E$&$0.54^{+0.00+0.12+0.06}_{-0.00-0.11-0.05}$&$0.80^{+0.06+0.12+0.06+0.07}_{-0.06-0.09-0.02-0.07}$\\

$\bar{B}^0 \rightarrow{D^0}(\rho^0\rightarrow){ \omega \pi^0}$ &$\frac{1}{\sqrt{2}}(E-C)$&$0.02^{+0.00+0.00+0.00}_{-0.00-0.00-0.00}$\\

$\bar{B}^0_s \rightarrow{D^+_s}(\rho^-\rightarrow){ \omega  \pi^-}$ &$T $&$0.68^{+0.00+0.14+0.07}_{-0.00-0.13-0.07}$&$0.88^{+0.05+0.07+0.00+0.06}_{-0.05-0.07-0.01-0.06}$\\

$  $ &$V_{cb}V_{us}^*$& $(\times 10^{-7})$\\

$\bar{B}^0_s \rightarrow{D^+}(\rho^-\rightarrow){\omega  \pi^-}$ &$E$&$ 1.21^{+0.11+0.00+0.16}_{-0.10-0.00-0.17} $\\

$\bar{B}^0_s \rightarrow{D^0}(\rho^0\rightarrow){ \omega  \pi^0}$ &$\frac{1}{\sqrt{2}}E$&$ 0.61^{+0.05+0+0.07}_{-0.05-0-0.06} $\\
 
 \hline 
 $ \bar{B} \rightarrow{\bar{D}}(\rho\rightarrow){\omega  \pi  } $ &$V_{ub}V_{cs}^*$& $(\times 10^{-7})$ \\
 
$B^- \rightarrow{D^-_s}(\rho^0\rightarrow){  \omega  \pi^0}$ &$\frac{1}{\sqrt{2}}T$&$ 11.05^{+0.00+2.32+0.45+0.87}_{-0.00-2.10-0.44-0.87} $\\

$\bar{B}^0 \rightarrow{D^-_s}(\rho^+\rightarrow){  \omega  \pi^+}$ &$T$&$ 20.21^{+0.00+4.24+0.82+1.59}_{-0.00-3.84-0.80-1.59} $\\

$\bar{B}^0_s \rightarrow{D^-}(\rho^+\rightarrow){  \omega  \pi^+}$ &$E$&$0.12^{+0.01+0.00+0.01+0.01}_{-0.01-0.00-0.01-0.01}  $\\

$\bar{B}^0_s \rightarrow{\bar{D}^0}(\rho^0\rightarrow){  \omega  \pi^0}$ &$\frac{1}{\sqrt{2}}E$&$0.06^{+0.01+0.00+0.01+0.00}_{-0.01-0.00-0.01-0.00} $\\

$  $ &$V_{ub}V_{cd}^*$&$(\times 10^{-7})$  \\
$B^- \rightarrow{D^-}(\rho^0\rightarrow){ \omega  \pi^0 }$ &$\frac{1}{\sqrt{2}}(T-A)$&$ 0.25^{+0.00+0.07+0.02+0.02}_{-0.00-0.06-0.02-0.02} $\\

$B^- \rightarrow{\bar{D}^0}(\rho^-\rightarrow){  \omega  \pi^-}$ &$C+A$&$ 0.29^{+0.01+0.04+0.02+0.02}_{-0.01-0.04-0.02-0.02} $\\

$\bar{B}^0 \rightarrow{D^-}(\rho^+\rightarrow){ \omega  \pi^+ }$ &$T+E$&$0.68^{+0.01+0.15+0.05+0.05}_{-0.01-0.14-0.05-0.05} $\\

$\bar{B}^0 \rightarrow{\bar{D}^0}(\rho^0\rightarrow){ \omega \pi^0}$ &$\frac{1}{\sqrt{2}}(E-C)$&$0.14^{+0.01+0.03+0.01+0.01}_{-0.01-0.03-0.01-0.01}$\\  
 \hline \hline
\end{tabular}
\end{center}
\end{table}

\begin{table}[tbhp]
\caption{Same as Table~\ref{rhoVP} but for the virtual effects of 
$\bar B_{(s)} \to D_{(s)} ( \omega \to)  \rho\, \pi $.}
 \label{omegarhopi}
 \begin{center}
\begin{tabular}{cccc}
 \hline \hline
~~~~Decay Modes ~~~~   &  ~~~~Amplitudes~~~~  & ~~~~$\mathcal{B}_{\text{FAT}}$~~~~\\

$\bar{B} \rightarrow{D}(\omega \rightarrow){\rho \pi}$ & $V_{cb}V_{ud}^*$ & \\

$\bar{B}^0 \rightarrow{D^0}(\omega\rightarrow){\rho^0 \pi^0}$ &$\frac{1}{\sqrt{2}}(E+C)$&$2.18^{+0.11+0.19+0.14}_{-0.11-0.38-0.13}\times 10^{-5}$\\



    & $V_{cb}V_{us}^*$ & \\

$\bar{B}^0_s \rightarrow{D^0}(\omega\rightarrow){\rho^0 \pi^0}$ &$\frac{1}{\sqrt{2}}E$&$5.36^{+0.48+0.00+0.58}_{-0.46-0.00-0.56}\times 10^{-8}$\\



\hline 
$ \bar{B} \rightarrow{\bar{D}}(\omega\rightarrow){\rho  \pi  } $ &$V_{ub}V_{cs}^*$&  \\
 
$B^- \rightarrow{D^-_s}( \omega\rightarrow){\rho^0 \pi^0}$ &$\frac{1}{\sqrt{2}}T$&$9.06^{+0.00+1.90+0.37+0.71}_{-0.00-1.72-0.36-0.71}\times 10^{-7}$\\

$\bar{B}^0_s \rightarrow{\bar{D}^0}(\omega\rightarrow){\rho^0 \pi^0}$ &$\frac{1}{\sqrt{2}}E$&$5.45^{+0.48+0.00+0.59+0.43}_{-0.46-0.00-0.56-0.43}\times 10^{-9}$\\

&$V_{ub}V_{cd}^*$&  \\

$B^- \rightarrow{D^-}(\omega\rightarrow){\rho^0 \pi^0 }$ &$\frac{1}{\sqrt{2}}(T+A)$&$5.69^{+0.00+0.93+0.38+0.45}_{-0.00-0.86-0.37-0.45}\times 10^{-8}$\\

$\bar{B}^0 \rightarrow{\bar{D}^0}(\omega\rightarrow){\rho^0 \pi^0}$ &$\frac{1}{\sqrt{2}}(E+C)$&$7.06^{+0.35+1.37+0.44+0.56}_{-0.35-1.24-0.43-0.56}\times 10^{-9}$\\  
                                                                            
 \hline \hline
\end{tabular}
\end{center}
\end{table}

\begin{table}[tbhp]
\caption{Same as Table~\ref{rhoVP} but for the virtual effects of 
$\bar B_{(s)} \rightarrow D (K^{*}\rightarrow)  \rho\, K$.}
 \label{KstarrhoK}
 \begin{center}
\begin{tabular}{cccc}
 \hline \hline
~~~~~~~~Decay Modes  ~~~~~~~  & ~~~~~~~~ Amplitudes  ~~~~~~~& ~~~~~~~~$\mathcal{B}_{\text{FAT}}$~~~~ \\

\hline

$ \bar{B} \rightarrow{D}(K^*\rightarrow){\rho K } $ &$V_{cb}V_{ud}^*$ \\


$\bar{B^0} \rightarrow{D^+_s}(K^{*-} \rightarrow){ \rho^- \bar{K}^{0}}$ &$E$
&$0.65^{+0.06+0.00+0.08}_{-0.06-0.00-0.08} \times 10^{-6} $\\

$\bar{B^0_s} \rightarrow{D^0}(K^{*0} \rightarrow){\rho^- K^+} $ &$C$
 & $1.16^{+0.04+0.24+0.08}_{-0.04-0.22-0.08}\times 10^{-5}$\\


$  $ &$V_{cb}V_{us}^*$  \\


$B^- \rightarrow{D^0}(K^{*-}\rightarrow){ \rho^- \bar{K}^0}$ &$T+C$
&$2.18^{+0.03+0.41+0.20}_{-0.03-0.37-0.19}\times 10^{-5}$\\


$\bar{B^0} \rightarrow{D^+}(K^{*-} \rightarrow){\rho^- \bar{K}^{0}}$ &$T$
& $1.64^{+0.00+0.34+0.17}_{-0.00-0.31-0.16}\times 10^{-5}$\\

$\bar{B^0} \rightarrow{D^0}(\bar{K}^{*0} \rightarrow){ \rho^+ K^-}$ &$C$
 & $0.79^{+0.02+0.17+0.05}_{-0.02-0.15-0.05}\times 10^{-6}$\\



$\bar{B^0_s} \rightarrow{D^+_s}(K^{*-} \rightarrow){\rho^- \bar{K}^{0}} $ &$T+E$
 & $1.75^{+0.01+0.38+0.18}_{-0.01-0.34-0.17}\times 10^{-5}$ \\
 
 \hline
 
 $ \bar{B} \to {\bar{D}}(K^*\to ){ \rho K } $ &$V_{ub}V_{cs}^*$&  \\


$B^- \rightarrow{\bar{D}^0}(K^{*-}\rightarrow){ \rho^- \bar{K}^0}$ &$C+A$
 & $2.66^{+0.08+0.45+0.18+0.21}_{-0.08-0.41-0.17-0.21}\times 10^{-7}$\\

$B^- \rightarrow{D^-}(\bar{K}^{*0} \rightarrow){\rho^+ K^-}$ &$A$
 & $2.36^{+0.00+0.00+0.16+0.19}_{-0.00-0.00-0.15-0.19}\times 10^{-8}$ \\


$\bar{B}^0 \rightarrow{\bar{D}^0}(\bar{K}^{*0} \rightarrow){\rho^+ K^-}$ &$C$
& $1.18^{+0.04+0.25+0.08+0.09}_{-0.04-0.22-0.08-0.09}\times 10^{-7}$ \\


$\bar{B^0_s} \rightarrow{D^-_s}(K^{*+} \rightarrow){\rho^+ K^{0}} $ &$T+E$
&$5.28^{+0.10+1.22+0.22+0.42}_{-0.11-1.09-0.22-0.42}\times 10^{-7}$
\\

$  $ &$V_{ub}V_{cd}^*$&  \\

$B^- \rightarrow{D^-_s}(K^{*0} \rightarrow){\rho^- K^+}$ &$A$
&$1.15^{+0.00+0.00+0.08+0.09}_{-0.00-0.00-0.08-0.09}\times 10^{-9}$\\



$\bar{B^0} \rightarrow{D^-_s}(K^{*+} \rightarrow){ \rho^+ K^{0}}$ &$E$
&$2.77^{+0.25+0.00+0.31+0.22}_{-0.24-0.00-0.29-0.22}\times 10^{-10}$\\


$\bar{B^0_s} \rightarrow{D^-}(K^{*+} \rightarrow){\rho^+ K^{0}} $ &$T$
&$2.52^{+0.00+0.53+0.17+0.20}_{-0.00-0.48-0.16-0.20}\times 10^{-8}$ \\

$\bar{B^0_s} \rightarrow{\bar{D}^0}(K^{*0} \rightarrow){\rho^- K^{+}} $ &$C$
&$4.94^{+0.37+1.04+0.33+0.39}_{-0.36-0.94-0.32-0.39}\times 10^{-9}$\\

  \hline \hline
\end{tabular}
\end{center}
\end{table}

\begin{table}[tbhp]
\caption{The virtual effects of $\bar B \to D(\rho, \omega \to) K\, K $ decays
in the FAT approach (${\mathcal B}^v_{ \rm {FAT}} $) [...]. }
 \label{Domega}
\begin{center}
\begin{tabular}{ccc}
 \hline \hline
~~~~~~~~~~~~~{Decay Modes} ~~~~~~~~~~~~~   &  ~~~~~~~~~~~~~${\mathcal B}^v_{ \rm {FAT}} $ ~~~~~~~~~~~~~   \\
\hline
 
 $B^-\to D^0 (\rho^-\to) K^- K^0$\;  &
        $6.59^{+0.10+1.21+0.61}_{-0.09-1.12-0.60}\times10^{-5}$ 
        \\
 $\bar B^0\to D^+ (\rho^-\to) K^- K^0$\;  &
            $4.54^{+0.03+0.99+0.48}_{-0.03-0.89-0.46}\times10^{-5}$ 
         \\ 
 $\bar B^0\to D^0 (\rho^0\to) K^+ K^-$\;    &
             $1.28^{+0.07+0.12+0.01}_{-0.07-0.24-0.01}\times10^{-6}$ 
          \\   
 $\bar B_s^0\to D_s^+ (\rho^-\to) K^- K^0$\; &
            $5.63^{+0+1.18+0.60}_{-0-1.07-0.60}\times10^{-5}$   \\
             $\bar B^0\to D^0 (\omega \to) K^+ K^-$\;  & 
       $1.54^{+0.08+0.13+0.02}_{-0.07-0.27-0.02} \times10^{-6}$ &
          \\    
   \hline \hline
\end{tabular}
\end{center}
\end{table}

The decay modes can be categorized based on 
the involved CKM matrix elements into three types:
Cabibbo-favored ($V_{cb}V^*_{ud}$), 
Cabibbo-suppressed ($V_{cb}V^*_{us}$), 
and doubly Cabibbo-suppressed ($V_{ub}V^*_{cs}$ 
and $V_{ub}V^*_{cd}$) processes.
The corresponding CKM matrix elements are listed 
in the second column of Tables~\ref{rhoVP} $-$ \ref{KstarrhoK}. 
This classification shows a clear hierarchy in branching fractions,
where the Cabibbo-favored modes are approximately $2 - 3$ orders 
larger than their doubly Cabibbo-suppressed counterparts 
within the same table. 
For decays governed by the same CKM matrix elements, 
the branching fractions also depend on the hierarchy of 
topological diagram amplitudes. 
Similar to the dynamics of the two-body $B$ decays ($B \to D V^*$),
the color-favored emission diagram ($T$) dominates in the 
the quasi-two-body decays ($B \to D (V^* \to ) VP$). 
Consequently, both the Cabibbo-favored and $T$ diagram dominated modes
are firstly to be measured experimentally. The CLEO Collaboration 
reported the first observation of $B^+ \to \bar D^{0} \omega \pi^+$ and 
 $B^0 \to \bar D^{-} \omega \pi^-$ with 
their total branching fractions to be  $(4.1 \pm 0.9 ) \times 10^{-3}$ 
and $(2.8 \pm 0.6 ) \times 10^{-3}$, respectively.
Furthermore, a spin parity analysis of the $\omega\, \pi^-$ pair
in the $D\, \omega \, \pi^-$ final state revealed a preference for 
a wide $1^-$ resonance~\cite{CLEO:2001sdm}. 
In our results, the virtual $\rho$ contributions to 
 $\bar{B}^0 \to D^+ (\rho^-\to) \omega \pi^-$
 and $\bar{B}_s^0 \to D_s^+ (\rho^-\to) \omega \pi^-$ modes 
 in Tab.~\ref{rhoVP}, are of nearly the same order as the 
 total branching fractions shown above. 
 Additionally, Ref.\cite{Belle:2015fvz} highlights a large $\omega\, \pi$ contribution
 from $\rho(770)$ in $B^0 \to \bar D^{*-} \omega \pi^-$,
 with a measured branching fraction  
$\mathcal{B} (B^0 \to \bar D^{*-}(\rho^+ \to ) \omega \pi^-) 
 =(1.48^{\, +0.37}_{\, -0.63}) \times 10^{-3}$.
 Given that $B \to D (\rho \to ) \omega \pi$ modes exhibit
 comparable virtual contribution to $B^0 \to \bar D^{*-} (\rho\to) \omega \pi^-$,
 they should also be measurable in experiments.
 
 The processes of $B \to D\, (\omega \to ) \rho\, \pi$ are listed in 
 Table~\ref{omegarhopi}.
 As the subprocesses of $\omega \to \rho^0 \pi^0$, $\omega \to \rho^- \pi^+$ and
 $\omega \to \rho^+ \pi^-$ satisfy the isospin symmetry,
 we only show the $B \to D\, (\omega \to ) \rho^0\, \pi^0$ in the table.
  For the decay $B \to D\, (\omega \to ) \rho\, \pi$, 
 the $\omega$ meson is produced in the initial $B$ decays, and 
 its subsequent decay to $\rho \pi$ would follow
 the Vector Meson Dominance mechanism proposed
 by GSW~\cite{Gell-Mann:1962hpq}. 
 While the branching fractions for this specific decay chain might not
 be measured, the data 
 $\mathcal{B} (\omega \to \pi^+ \pi^- \pi^0)=(89.2 \pm 0.7)\%$~\cite{ParticleDataGroup:2024cfk},
 suggests that the subprocess $\omega \to \rho \pi$ is likely significant.
Its branching fraction accounts for about $2/3$ of the total decay ratio of $ \omega \to 3 \pi$~\cite{Cabrera:2013zga},
 and the contribution ratio even reach to $80\%$~\cite{Lucio:1999mb}.
 The theoretical estimates for the decay rate for  $\bar B_{(s)} \to D_{(s)} \, (\omega \to ) \rho\, \pi$ 
 can  utilize the narrow width approximation for the $\omega$ resonance,
 expressed as 
 $\mathcal{B}(B \to D\, \omega) \, \times\, \mathcal{B}(\omega \to \rho \pi)$,
 where $\mathcal{B}(\omega \to \rho \pi)$ is 
 effectively $1/2$ of $\mathcal{B} (\omega \to \pi^+ \pi^- \pi^0)$ without considering
  the interference of the respective three $\rho$ meson~\cite{Lucio:1999mb,Cabrera:2013zga}.
Taking $\bar B^0 \to D^0 (\omega \to) \rho \pi $ as an example, 
its branching fraction is given by
 \begin{align}
 \label{omega3pi}
 &\mathcal{B}(\bar B^0 \to D^0 \, \omega) \, \times\, \mathcal{B}(\omega \to\rho \pi)\nonumber\\
&\approx  \mathcal{B}(\bar B^0 \to D^0 \, \omega) \, \times\, \frac{1}{2}\, \mathcal{B}(\omega \to 3 \pi)
\nonumber\\
&=(1.16 \pm 0.25)\times 10^{-4}\, ,
 \end{align}
 where two-body decays is calculated to be 
 $\mathcal{B}(\bar B^0 \to D^0 \, \omega)=(2.59 \pm 0.55) \times 10^{-4}$.
The value in Eq.(\ref{omega3pi}) is approximately 
three times of the branching fraction 
$\mathcal{B}(\bar B^0 \to D^0 \, (\omega \to) \rho^0 \pi^0) $ 
listed in Table~\ref{omegarhopi}, e.g., ($0.65^{+0.08}_{-0.13}) \times 10^{-4}$,
by summing all $\rho \, \pi$ sub-channels,
$\omega \to \rho^0 \pi^0$, $\omega \to \rho^- \pi^+$ and
 $\omega \to \rho^+ \pi^-$.
 This consistency serves as a crosscheck of 
 the theoretical framework of FAT applied in this analysis.
 Experimental confirmation would require reconstructing the 
$\omega \to \pi^+ \pi^- \pi^0$  and the $ \rho \to \pi \pi$,
which is challenging due to the neutral pions but feasible 
with high statistics datasets in Belle II and LHCb.
 
The $B \to D\, (K^* \to ) \rho\, K$ decays are represented 
in Table~\ref{KstarrhoK}.
Due to isospin symmetry of strong decays, 
the subprocess of $B \to D\, (K^* \to ) \rho\, K$ decays satisfy the following relationship,
\begin{eqnarray}
 \mathcal{B}(\bar{K}^{*0} \to K^- \pi^+) &=& 2\, \mathcal{B} (\bar{K}^{*0} \to \bar{K}^0 \pi^0),\\
 \mathcal{B} (K^{*-} \to \bar{K}^0 \pi^-)&=& 2\, \mathcal{B}(K^{*-} \to K^- \pi^0)\, ,
 \end{eqnarray}
 where  
$ \bar{K}^{*0} \to  K^- \pi^+$ and $ K^{*-} \to  K^- \pi^0$
 happen through sea quark pair $u \bar u$,
and $ \bar{K}^{*0} \to  \bar{K}^0 \pi^0$ 
and $ K^{*-} \to  \bar{K}^0 \pi^-$ via $d \bar d$ pair.
Thus, modes like $B \to D\, (\bar{K}^{*0}\to) \bar{K}^0 \pi^0$ and 
 $B \to D\, (K^{*-} \to ) K^- \pi^0$ need not be listed in the 
 Table~\ref{KstarrhoK} separately.
In contrast with $\omega$ resonance, the decay $K^* \to  \rho\, K$
is not a dominant mode. Instead, $K^*$ typically decays via
$K \pi$, e.g., $\mathcal{B}(K^{*0} \to  K^+ \pi^-) \sim 100\%$.
Hence no experimental data is available for $K^* \to \rho \pi$, 
and only the upper limits can be given for $K^{*-} \to K^- \pi^+ \pi^-$,
$K^{*-} \to K^0 \pi^- \pi^0$ and 
$K^{*0} \to K^0 \pi^+ \pi^-$~\cite{Amsterdam-CERN-Nijmegen-Oxford:1977ujs}.
However, as noted for the $\rho$ resonance, they also decay to $\pi \pi$ with 
$\mathcal{B}(\rho\to  \pi \pi) \sim 100\%$, while their off-shell effects 
dominate in $ D \omega \pi$ final state.
Following the observation of $B^+ \to \bar D^{0} \omega \pi^+$ and 
 $B^0 \to \bar D^{-} \omega \pi^-$ by CLEO Collaboration,
we predict that the decays $B^- \rightarrow{D^0}(K^{*-}\rightarrow){  \rho^0 \, K^-}$ and 
$\bar B^0 \rightarrow{D^+}(K^{*-}\rightarrow){  \rho^0\, K^-}$ with 
 branching fraction of $\mathcal{O}(10^{-5})$ are also expected to 
 be observed in future experiments. 
 Searches for more exotic intermediate resonance decays of $K^* \to \rho K$
 would require dedicated analyses by future high-statistics experiments,
such as Belle II and LHCb.

In Tab.~\ref{Domega}, we present the virtual effects
of the Cabibbo-favored modes $\bar B \to D(\rho, \, \omega \to) K K$, 
as we calculated in Ref.~\cite{Zhou:2024qmm}, for comparison.
A notable observation is that the off-shell effects of 
$\rho, \omega$ resonance in $\bar B \to D(\rho \to) \omega \pi$ and 
 $\bar B \to D(\omega \to) \rho \pi$ 
are approximately two orders of magnitude larger than those in $\rho \to K\, K$,
$\omega \to K\, K$, respectively.
This significant difference arises primarily from the much stronger 
coupling strength of $\rho \to \omega \pi$ and $\omega  \to \rho \pi$ 
($g_{\rho \omega \pi} = 16.0 \pm 2.0$\, $\text{GeV}^{-1}$,
$g_ \omega {\rho \pi} = 16.0 \pm 2.0$\, $\text{GeV}^{-1}$)
compared to that of $\rho \to K K$ and  $\omega \to K K$ 
($g_{\rho KK} = 3.2$, $g_{\omega KK} = 3.2$~\cite{Zhou:2024qmm}).
These strong couplings can be calculated by some 
nonperturbative methods, such as QCD sum rules,
so as to understand the strong decay mechanism of $\rho, \omega$ mesons.


\subsection{Comparison with the results in the PQCD approach}
Since most quasi-two-body decays 
$B_{(s)} \rightarrow D_{(s)} (V^* \rightarrow)P V$ 
have not yet been measured experimentally, 
we compare our results with the PQCD predictions 
from Ref.~\cite{Ren_2024}, listed in the last column of
 Table~\ref{rhoVP} for comparison. 
Our calculation for $B^- \to D^0 (\rho^- \to ) \omega \pi^-$,
$\bar B^0 \to D^+ (\rho^- \to ) \omega \pi^-$,
and $\bar B_s^0 \to D_s^+ (\rho^- \to ) \omega \pi^-$
are consistent with the PQCD calculations but are
in fact more precise.
The PQCD results mainly includes some hadronic uncertainties,
such as shape parameters $\omega_B$ and Gegenbauer moments,
which are comparable to the uncertainties arising from 
form factors and decay constants in our analysis.
However, it does not account for 
uncertainties from $1/m_b$ 
power corrections such as annihilation and hard-scattering contributions,
 and higher order $\alpha_s$ effects, 
which would introduce significant additional uncertainties 
in their results predictions, the same as in two-body decays 
$B \to D V^*$~\cite{Li:2008ts}.

Actually, in the PQCD approach, the light-cone 
distribution amplitude (LCDA) of the $\omega \pi$ 
meson pair originating from the P-wave resonance 
$\rho$ can be expressed in terms of time-like form 
factors $F_{\omega \pi}$, parameterized using 
the RBW distribution. 
This treatment of resonant decays like $\rho \to \omega \pi$
is fully compatible with the framework of the FAT approach. 
The key distinction between the two methods lies in 
their handing of the weak decay amplitudes for 
$B \to D V^*$.
As discussed in Sec.~\ref{Introduction}, the FAT approach 
determines the nonperturbative parameters 
associated with the topological diagrams ($C$, $E$ and $A$)
by a global fit to all experimental data of $B \to D V^*$.
This data-driven constraint ensures that these parameters 
are tightly controlled. In contrast, PQCD face large 
uncertainties from non-perturbative parameters 
as the $1/m_b$ power corrections
and high order $\alpha_s$ effects are hard to be 
calculated under the PQCD framework.

Meanwhile, the hierarchy of topological amplitudes 
differs between the two approaches. 
In the FAT framework, the magnitude 
of $C$ diagram amplitude is larger than that of $E$,
 i.e.,$\left|C \right| > \left|E\right|$, as shown in Eq.(\ref{fitpara}).
 In contrast, PQCD calculations~\cite{Li:2008ts} yield
  $\left|C \right| \sim\left|E\right|$ due to
the absence of power corrections and higher-order 
radiative contributions.
It is well established that the $T$ diagram remains 
factorizable to all orders in  $\alpha_s$ for these decays,
making the perturbative calculation reliable. 
Accordingly, our results of the $T$ diagram 
dominated decay modes (listed in the top of Tables \ref{rhoVP})
align with PQCD predictions. 
However, for decays governed by $C$ amplitude, 
especially for those with branching ratio of $\mathcal{O} (10^{-5})$, 
are waiting  to be tested by the future experiments and 
PQCD calculations. 


 \section{Conclusion}\label{sec:4}
Motivated by the searches for virtual effects of intermediate 
resonances, we present a systematic analysis of the quasi-two-body 
decays $B_{(s)} \to D_{(s)} (V^* \to) V P$ where the intermediate 
ground states $V^*$ are $\rho,\, \omega, \, K^*$. These decays
proceed via $b \rightarrow \,c\, $ or $b \rightarrow \,u\, $ transitions,
corresponding to $\bar B_{(s)} \to D_{(s)} V^*$ and 
$\bar B_{(s)} \to \bar D_{(s)} V^*$, respectively.
The resonant states $V^*$ subsequently decay into $VP$ pairs 
 through strong interactions. Under the framework of FAT,
 we incorporate all nonperturbative contributions in 
 $B_{(s)} \to D_{(s)} V^*$ to achieve a precise description 
 of the first subprocess. For the intermediate resonances 
 decays, we apply the RBW function as usually used in 
experiments and the PQCD approach. 

We categorize $B_{(s)} \to D_{(s)} (V^* \to) V P$ into 
three groups according to the intermediate resonances, 
e.g., $B_{(s)} \to D_{(s)} (\rho \to)  \omega \pi $, 
 $B_{(s)} \to D_{(s)} (\omega \to) \rho \pi $ 
and $B_{(s)} \to D_{(s)} (K^* \to) \rho K$.
Since the pole masses of $\rho,\, \omega, \, K^*$
lie below the invariant mass threshold of their 
corresponding final-state pairs, these resonances 
can only contribute through their BWT distribution. 
We calculate the branching fractions for all  three 
decay types. Our analysis shows that the virtual 
contributions of $\rho,\, \omega, \, K^*$  
play a crucial role in the three-body decays
$B_{(s)} \to D_{(s)} V P$. Some modes exhibit 
decay rates up to $\mathcal{O}(10^{-4})$,
comparable to their total branching fractions.
 For decays involving $\omega \to \rho \pi$,
our results are in good agreement with the 
experimental analysis of $\omega \to \pi^+ \pi^- \pi^0$.
In particular, for $B^- \to D^0(\rho^-\to) \omega \pi^-$, 
$\bar{B}^0 \to D^+ (\rho^-\to) \omega \pi^-$, and 
$\bar{B}^0_s \to {D^+_s}(\rho^-\to){ \omega  \pi^-}$
previously calculated in PQCD approach, 
our predictions agree with them but with smaller 
uncertainties. Since most quasi-two-body decays 
$B_{(s)} \rightarrow D_{(s)} (V^* \rightarrow)P V$ 
remain unmeasured experimentally.
Our predictions for both the Cabibbo-favored
 and $T$ diagram dominated modes 
with branching fractions of order $10^{-6}-10^{-4}$  
are promising targets for future high statistics experiments, 
such as in Belle II and LHCb.


\section*{Acknowledgments}
The work is supported by the National Natural Science Foundation of China
under Grants No.12465017 and No.12075126.

\bibliographystyle{bibstyle}
\bibliography{refs}

\end{document}